# Liquid crystalline cellulose-based nematogels

*Qingkun Liu[1], Ivan I. Smalyukh[1,2,3]\**

[1] Department of Physics, University of Colorado, Boulder, CO 80309, USA
[2] Department of Electrical, Computer, and Energy Engineering, Materials Science and Engineering Program, and Soft Materials Research Center, University of Colorado, Boulder, CO 80309, USA;
[3] Renewable and Sustainable Energy Institute, National Renewable Energy Laboratory and University of Colorado, Boulder, CO 80309, USA

\*Corresponding author. Email: ivan.smalyukh@colorado.edu

**Summary:** We introduce electrically tunable gels formed by an ordered network of cellulose nanofibers infiltrated with a liquid crystal.

## Abstract

Physical properties of composite materials can be pre-engineered by controlling their structure and composition at the mesoscale. Yet, approaches for achieving this are limited and rarely scalable. We introduce a new breed of self-assembled nematogels formed by an orientationally ordered network of thin cellulose nanofibers infiltrated with a thermotropic nematic fluid. The interplay of orientational ordering within the nematic network and that of the small-molecule liquid crystal around it yields a composite with highly tunable optical properties. By means of combining experimental characterization and modeling, we demonstrate sub-millisecond electric switching of transparency and facile responses of the composite to temperature changes. Finally, we discuss a host of potential technological uses of these self-assembled nematogel composites, ranging from smart and privacy windows to novel flexible displays.

## Introduction

Composite materials exhibit physical properties that depend on their structure and composition at the mesoscale. The approaches for controlling the mesoscopic structure and the ensuing properties are, however, limited and rarely scalable. This is especially true for a broad family of materials, called "gels", which finds important technological and scientific applications, ranging from biomedical uses of hydrogels to applications of aerogels in thermal insulation and detection of Cherenkov radiation (1-3). Despite of the large diversity, these materials share a common feature of having three-dimensional (3D) networks of nanoparticles or molecules surrounded by a fluid- or a gas-like medium. These networks are often anisotropic in nature, made of rod-like or ribbon-like and other kinds of anisotropic building blocks (4), and sol-gel transitions are often accompanied by various types of liquid crystal (LC) phase behavior (5). For example, cellulose-based gels are typically composed of thin nanofibers, which can be organized in an orderly fashion while being part of the 3D network (6). Typically only isotropic fluids are used in gel materials, such as water in hydrogels and alcohols in organogels (7,8), though there have been studies of how silica gel confinement affects the ordering of different LCs (9). On the other hand, in related classes of soft materials, such as polymer-dispersed and polymer-stabilized LCs, the orientational ordering of the mesophase is mechanically coupled to the polymer matrix or network, which can be anisotropic (9). However, the means of robust control of the composite properties in these systems remain limited and the challenges of scalable cost-effective manufacturing hinder a host of potential technological uses.

In this work, we introduce a new breed of composite materials that combine ordered nematic LC gel networks based on cellulose nanofibers and a small-molecule thermotropic LC fluid similar to that used in LC displays. We pre-engineer the interplay of LC gel network and the LC fluid surrounding it to enable new composite properties that, to the best of our knowledge, cannot be achieved otherwise. We demonstrate that such composites can be switched by applying electric fields or by varying temperature, or by a combination of these two external stimuli, with the sub-millisecond response times and switching characteristics consistent with a simple model based on elasticity and surface anchoring properties of LCs. We characterize the structure and both mechanical and optical properties of the ensuing flexible nematogel films, showing that scalable fabrication of such composite materials can lead to their technological uses and application as diverse as smart and privacy windows, electro-optic shutters and polarizer-free information displays. Considering the scientific and technological impacts exhibited by distantly related classes of mesostructured soft materials, such as polymer-dispersed and polymer-stabilized LCs (9-13) and LC elastomers (16,17), we foresee that our gels will find a large number of practical applications and will reveal a host of new fundamental phenomena arising from the interplay of LC ordering and tunable mesoscale networks of cellulose and related systems.

**Results**

**Liquid crystalline phase behavior of nematogels**

The cellulose nanofibers (CNFs) (Fig. 1A,E) were prepared by employing the method of 2,2,6,6-tetramethylpiperidine-1-oxyl radical (TEMPO)-mediated oxidation of native cellulose. During this process, the C6 primary hydroxyl group of the β-1,4-linked D-glucose units is partially converted to C6 carboxylate group, as shown in the inset of Fig.1A (here, according to the common description of the chemical structure of cellulose, C6 is the carbon atom which is linked with primary hydroxyl group and is not positioned in the ring of the chair conformation). The electrostatic charging of the carboxylate anion on the surface of CNFs provides stabilization of the LC colloidal dispersions against aggregation in polar solvents like water. CNFs readily form a nematic lyotropic LC when dispersed in water at concentrations above the critical Onsager concentration. The high length-to-width aspect ratio of CNFs (typically within 100-300) assures the emergence of nematic LC phase behavior at vanishingly low volume fractions <1%, which can be well understood on the basis of the Onsager theory while also accounting for surface charging, and nanometer-range Debye screening length (18,19). Similar to other lyotropic nematic systems (20), the LC of CNFs can be aligned by shearing, which transforms a multi-domain sample with defects into an aligned monodomain LC sample with the relatively uniform shearing-defined director $N_s$, though small variations of local ordering of the nanofibers still remain (just like in other shear-aligned lyotropic LC materials). Cross-linking of these nanofibers (Fig. 1B,F-H) by hydrogen bonds between carboxyl groups after adding acid (see details in the methods section) transforms the colloidal LC fluid into a hydrogel while preserving its ordered structural features and preferential ordering of the cellulose nanofibrils along $\mathbf{N}_s$. Sequential exchange of the fluid within the gel, which includes the replacement of water with isopropanol and then with the nematic LC 4-cyano-4'-pentylbiphenyl (5CB), transforms this gel solid first into an ordered organogel and then into a nematogel (Fig. 1C,D,G,H), as we describe in details in the experimental methods section below.

The phase behavior of the nematogel as a composite is substantially different from that of the pristine bulk 5CB LC. This new behavior is caused by paranematic ordering of 5CB molecules induced by the network of nanofibrils with surfaces prompting their tangential orientation with respect to the nanofibers, similar to namatic LCs in other confinement geometries (9-13). The thermodynamic phase behavior of condensed matter is often modified by external fields and confinement so that material systems can exhibit ordered states even at temperatures at which they cannot be thermodynamically stable without fields or nanoconfinement (21). In the case of nematic LCs, the paranematic state is such a state that typically can exist in presence of fields or confinement even at temperatures above the nematic-isotropic phase transition temperature. For example, such effects of confinement-induced paranematic ordering have been studied previously in polymer-dispersed LCs and other nanoscale confinement geometries by Doane and others (10-12). Optical observations, including light transmission and polarizing microscopy textures, reveal that the nematogel composite formed by 5CB within the CNF network exhibits continuous nematic-like behavior as the sample is heated above the temperature of order-disorder transition of pristine 5CB. Instead of disappearing at the temperature of $T_{NI} \approx 35.3$ °C of the nematic-isotropic transition of pure bulk 5CB samples, the long-range ordering of 5CB molecules and director (manifested by the strong birefringence and light scattering) persists up to $T_{PNI} \approx 38.5$ °C of paranematic-isotropic transition of this composite system. Short-range paranematic ordering near cellulose nanofibers persists even above this transition, though birefringence and scattering drop down abruptly at $T_{PNI} \approx 38.5$ °C, which is similar to LCs in other nanoscale confinement geometries (9-12). Although the birefringence drops down at the paranematic-isotropic transition and gradually reduces further with increasing temperature, the composite stays weakly birefringent even at high temperatures of about 90 °C, which is due to the orientational ordering of cellulose nanofibers within the gel network itself.

The flexible nematogel film derives its unusual physical properties from an interplay of orientational ordering within the cross-linked nematic gel network of CNFs and the nematic fluid infiltrating it (Fig. 2), where the long-range ordering direction of 5CB molecules tends to follow $\mathbf{N}_s$. In polydomain nematogel samples obtained with weak or no shearing, the average ordering directions of the cellulose nanofibrils are coupled to the nematic director, as well as to the director of a paranematic state of infiltrated LC induced by the cellulose network in the pre-transition region, which can be examined using a polarizing optical microscope (POM) (Fig. 2).

**Thermally-responsive scattering of nematogels**

An interesting open question relates to how properties of nanocellulose-based gels can be modified by loading their mesoporous structure with stimuli-responsive materials, which can be controlled by temperature, fields, and other external stimuli. To explore this, we investigate nematogels with two kinds of the CNF network: (1) one with 5.6 wt.% of CNFs with a smaller-aspect ratio and dimensions 7 nm×400 nm (CNF1, Fig. 2A) and (2) one with 0.5 wt.% of CNFs of larger aspect ratio and length 4.8 nm×1.2 μm (CNF2, Fig. 2D), though it should be noted that in the process of gel formation the CNFs can form bundles (Fig. 1F-H) with diameters considerably larger than the ones in the initial LC colloidal dispersion (Fig. 1E). These nematogels are both infiltrated with the thermotropic LC 5CB and exhibit reversible switching of transparency. The nematogel films with unaligned CNFs appear "milky" (Fig. 2A,D) at room temperature, but become fairly transparent when heated several degrees above the temperature of nematic-isotropic transition $T_{NI}$ of the pristine 5CB nematic fluid and just above $T_{PNI}$ of the nematogel composite (Fig. 2B,E). The CNFs network preserves its nematic-like order and shows birefringence under POM at the temperatures not only above $T_{NI}$ of 5CB but even above $T_{PNI}$ of the nematogel composite (Fig. 2C,F). Even above $T_{PNI}$, optical anisotropy due to the nematic network of cellulose nanofibers is additionally enhanced by the short-range paranematic ordering induced by the CNF-5CB interfaces (19), as already discussed above.

At temperatures above $T_{PNI}$, the nematogel exhibits high transparency across a wide range of the visible and near infrared spectrum, even higher than that of the original cellulose-based hydrogel or organogel infiltrated with ethanol (Fig. 2J). The high transmission is enabled by the effective matching of refractive index of CNFs and the 5CB fluid at these elevated temperatures. The CNFs are optically anisotropic, containing both crystalline and amorphous regions. The crystal structure of the crystalline component of CNFs is reminiscent to that of elementary fibrils found in the cellulose derived from wood (22). Both of the studied types of nanofibers consist of 36 cellulose chains arranged in monoclinic Iβ crystal structure with a P2$_1$ space group, containing both amorphous and crystalline regions (22). The refractive index of CNF with ideally oriented cellulose I crystal structure was previously reported to be $n_e^{CNF}=1.618$ for normally incident light polarized along the fiber and $n_o^{CNF}=1.544$ for the linear polarization in the transverse direction (23), though these values depend on the degree of crystallinity, where $n_e$ and $n_o$ are the extraordinary and ordinary refractive indices, respectively. However, due to the combination of imperfect chain orientations and the existence of amorphous regions, CNFs show weaker optical anisotropy and extraordinary and ordinary refractive indices closer to each other $n_e^{CNF} \approx 1.59$ and $n_o^{CNF} \approx 1.53$, different from the theoretical values estimated for ideal crystals without the amorphous content (23). In the nematic phase, 5CB is also a positively anisotropic uniaxial optical material with the extraordinary and ordinary refractive indices $n_e^{5CB}=1.698$ and $n_o^{5CB}=1.535$ at 589 nm and 24°C (24). Upon heating the pristine 5CB above $T_{NI}$ (or $T_{NI}$ for our composites), this material becomes an isotropic fluid with a single polarization-independent refractive index $n_{iso}^{5CB}=1.586$ at 589 nm and temperatures 39 °C and higher (24), though even this value is altered by the cellulose network within the composite. The surfaces of CNFs impose tangential alignment for the director of the small-molecule nematic 5CB. Because of these boundary conditions, orientational ordering within 5CB mechanically couples to that of the CNF network composed of fibrils oriented, on average, parallel to each other along **N**$_s$ and to the nematic director of 5CB infiltrating it. This coupling in alignment can be seen by examining domains of the CNF-5CB composite at a temperature right above paranematic-isotropic transition (when the long-range ordering disappears and the nematic-like organization persists only in small regions near the nanofibril surfaces) between crossed polarizers (Fig. 2C,F). For example, for the linearly polarized incident light with the polarization direction along the director of the CNF-5CB composite, because of the strong mismatch of refractive index ($n_e^{5CB} - n_e^{CNF} \approx 0.1$) between CNFs and 5CB in the nematic phase, the scattering is large. On the other hand, for the temperature $T>T_{PNI}$, the mismatch of the refractive index ($n_e^{CNF} - n_{iso}^{5CB} < 0.01$, with the gradual changes of the index of 5CB on the nanoscale in the vicinity of nanofibers due to the surface-induced paranematic ordering) is much smaller, causing substantially less scattering of light. This analysis, with a similar conclusion, can be extended to unpolarized light. Beyond the CNF-5CB index mismatch in the nematic state of the nematogel composite, individual domains of the composite can have misoriented ordering directions or even defects, which yield additional spatial variations of the effective refractive index within 5CB. In thick samples, director fluctuations in the nematic state of 5CB also contribute to scattering and hazy appearance below $T_{PNI}$. The large variety of sources of scattering in the nematic state of 5CB within the composite causes hazy appearance of the composites across the entire visible part of spectrum. On the other hand, the scattering is mainly of Rayleigh type above $T_{PNI}$, arising mainly just from the mismatches of indices between CNFs and isotropic 5CB, consistent with the much stronger scattering of the violet-blue light as compared to red light.

A flexible device with thermally and electrically responsive light transmission can be fabricated by sandwiching the nematogel into a plastic LC cell with transparent indium tin oxide (ITO) electrodes (Fig. 2G). At room temperature, the cell is in an opaque state because of the scattering of visible light discussed above, but it becomes highly transparent at an elevated temperature $T>T_{PNI}$ because of the closer CNF-5CB refractive index matching in the isotropic phase of 5CB (Fig. 2H-L). For example, the relative visible-spectrum-averaged light transmission $\eta_{trans}$ of a 30 μm-thick LC cell with the confined nematogel can be increased from 5% to 97% in response to changing temperature from 32 °C to 41 °C (Fig. 2K), where the transmittance $\eta_{trans}$ is defined as the ratio of direct transmitted and incident light intensities, with the reflections at the gel-air interfaces and the absorption of plastic substrate subtracted. The temperature-dependent light transmission (Fig. 2K) further confirms the presence of the CNF network-induced paranematic ordering of 5CB, with the paranematic-to-isotropic transition temperature $T_{PNI} \approx 38.5°C$ of the nematogel composite, which is significantly higher than the $T_{NI}$ of the pristine 5CB (35.3°C). This behavior can be further understood by invoking the theoretical models describing the behavior of LCs confined in nano-porous confined structures, as we detail below (25-26).

**Sub-millisecond electric switching**

The transparency of the nematogel composites can be controlled by electric fields. When the polarization of incident light is parallel to the director $\mathbf{N}_s$ of the CNF-5CB composite at no applied electric field, with the rod-like 5CB molecules following the fiber orientations, the mismatch of the refractive index between 5CB and CNFs is $n_e^{5CB} - n_e^{CNF} \approx 0.1$, so the nematogel scatters strongly. The scattering leads to low light transmittance and to the opaqueness of the nematogel cells. When we apply an AC sinusoidal voltage within $f$=1-50 kHz to the nematogel (note that high-frequency electric fields are needed to demonstrate the fast response of nematogels that we discuss below), the 5CB molecules reorient along the electric field, so the mismatch of refractive index between 5CB and CNFs constitutes $n_e^{CNF} - n_o^{5CB} < 0.06$ at 24°C (and <0.03 when the temperature is close to $T_{PNI}$) (24), smaller than the value before applying voltage. Additionally, the nematic director domains of spatially varying director orientation and refractive index disappear as the LC as aligned with the electric field orientation and even the light scattering due to director fluctuations is suppressed in an applied field. Consequently, electric field can switch the light transmission from 18% to 70% at 600 nm (Fig. 2J). By measuring the light transmittance of the cells as a function of the applied voltage, we determine the electrooptic switching characteristics and the critical field of the Freedericksz transition $E_c$ at different temperatures (Fig. 2J-L). The threshold electric fields remain practically unchanged at ≈3.0 V/µm at the studied temperatures, though transmission of both the "on" and "off" states tend to be higher at elevated temperatures (Fig. 2L), which can be attributed to the effects related to matching of refractive indices of 5CB and CNF at different temperatures. The voltages needed for the nematogel switching depend on the thickness of the film and can range from ~10V for thin films in the micrometer range to over 100 V for thick films with thickness of 100 µm and higher.

To understand the mechanical coupling between the gel network of CNFs and the director of 5CB, as well as to get insights into the switching behavior of nematogels (Fig. 3), we apply a physical model accounting for the LC's elastic, surface anchoring and dielectric properties in the used geometry of cellulose-based nematogels. In particular, we adopt (and extend to our system) a model that describes electric switching of nanoscale-dispersed LC composites developed previously for polymer dispersed LC systems (27). We assume that the LC is compartmentalized in domains with a periodicity of $a$ along the $x$ axis and $b$ along the $y$ axis encircled by CNFs, as illustrated in the inset of Fig. 3A. By assuming that the CNF nanofibrils are perfectly straight and very long as compared to their width, we can treat the problem as two dimensional in nature while assuming that the structure is translationally invariant along the $z$-axis of the coordinate system introduced in Fig. 3A. We use the one-elastic-constant approximation within the Frank-Oseen theory of orientational elasticity (19,28) and model the geometry of domains of the nematic director in the field-on state by rectangles of dimension $a$ along the $x$-axis and $b$ along the $y$-axis. Minimization of the total free energy, including its elastic and electric field coupling bulk terms and the surface anchoring terms due to the weak anchoring conditions at the surfaces of the CNF network, yields the equilibrium director configurations within the nematogel at different fields, albeit the high-field behavior cannot be easily accounted for analytically (19,28) and requires numerical studies. We therefore first consider this problem at the onset the Freedericksz transition, at the lowest threshold field when the dielectric torque overcomes the elastic and surface anchoring torques in prompting the director realignment (19,28). This problem can be treated for the LC nematogel cell of gap thickness $d$ along the applied field direction, assuming that it can then be extended to understand the realignment transition in each of the identical rectangular nematogel domains confined by the CNF network. The total bulk free energy density (per unit volume within each compartment) of the nematic LC can be expressed as

$$f = \frac{K}{2}\left[\left(\frac{\partial \theta}{\partial x}\right)^2 + \left(\frac{\partial \theta}{\partial y}\right)^2\right] - \frac{1}{2}\varepsilon_0 \Delta\varepsilon E^2 \sin^2\theta \tag{1}$$

where $\theta(x,y)$ is the distortion angle of the nematic director with respect to the $z$ axis, $K$ is an average Frank elastic constant of the LC, $\varepsilon_0$ is the vacuum permittivity, $\Delta\varepsilon$ is the LC dielectric anisotropy, $E$ is the electric field (27). With finite surface anchoring on the perimeter of each of the rectangular director domains, the geometric parameters $a$ and $b$ have to be modified to account for the finite-strength boundary conditions at the CNF-5CB interfaces. For this, we assume that the boundary conditions are tangentially degenerate and that the surface anchoring energy per unit area can be expressed in the Rapini–Papoular form: $f_s = W \sin^2\theta/2$, where $W$ is the polar surface anchoring strength coefficient characterizing director-CNF coupling at the surfaces in the corners of the rectangular domains. The average distance between the centers of the randomly distributed cellulose fibrils changes with the CNF volume fraction (dimensionless, or can be expressed in %) $c$ as $d' = \alpha/\sqrt{c} - D$, where $D$ is the diameter of CNFs and $\alpha$ is a geometry-dependent coefficient with units of length. Considering unidirectional aligned of infinitely long fibers assumed in our simplified geometry model, the volume fraction varies as $c = [D/(d'+D)]^2$, so that we can roughly estimate $d'$ as $d' \approx D/\sqrt{c} - D$. The parameters $a$ and $b$ vary with increasing concentration as $1/a = 1/d + 1/d'$ and $1/b = 1/d'$, where the characteristic length $d'$ is determined by the details of the CNF network geometry discussed above. The critical threshold field needed for director realignment in our nematogel system with finite anchoring can be then expressed as (27)

$$E_c = \sqrt{\frac{K\pi^2}{\varepsilon_0 \Delta\varepsilon}\left[\left(\frac{1}{d} + \frac{1}{d'+2K/W}\right)^2 + \left(\frac{1}{d'+2K/W}\right)^2\right]}, \tag{2}$$

which, as we show below, is consistent with experimentally determined parameters. In analogy with the polymer dispersed LCs (27), we can then model the anticipated electrically driven director realignment response times. Since the falling time is $\tau_{falling} E_c^2 = \gamma/(\varepsilon_0 \Delta\varepsilon)$, the falling and rising response times can be explicitly expressed as (27)

$$\tau_{falling} = \frac{\gamma}{K\pi^2}\left[\left(\frac{1}{d} + \frac{1}{d'+2K/W}\right)^2 + \left(\frac{1}{d'+2K/W}\right)^2\right]^{-1} \tag{3}$$

$$\tau_{rising} = \frac{\tau_{falling}}{(E/E_c)^2 - 1} \tag{4}$$

where $\gamma$ is the rotational viscosity of 5CB infiltrating the gel.

In the studied composite material, the concentration weight fraction of CNF is 0.19 wt.%, which translates to the volume fraction of 0.127 vol.% when using the density of cellulose $\rho_{CNF} = 1.5$ g/cm$^3$. Taking the experimental diameter $D = 4.8$ nm of CNFs (Fig. 1E-H) and assuming that the CNF network has homogeneous alignment due to shearing and also that the fibrils do not touch each other, we calculate average distance $d'$ between the CNFs to be 130 nm, which is consistent with dimensions of the CNF skeleton and morphology revealed by scanning electron microscopy (SEM) displayed in Fig.

1G. By substituting the average elastic constant $K \approx 5$ pN and dielectric anisotropy $\Delta\varepsilon=11.25$ of 5CB at $f=50$ kHz (29), $\varepsilon_0 =8.85\times10^{-12}$ F/m, $d=30$ μm and the experimental value $E_c=3.0$ V/μm (Fig. 2L) into Equation 2, we obtain an estimate of the surface anchoring coefficient $W=4.9\times10^{-5}$ J/m$^2$. This estimate is consistent with the weak surface anchoring boundary conditions assumed in the model, as well as is close to the independent measurements of polar surface anchoring coefficient at the cellulose-5CB interfaces reported in literature (30,31).

The nematogel exhibits ultra-fast electric switching of transparency, which is of interest for a broad range of technological uses, including such applications as optical shutters, electro-optic modulators, privacy windows and transflective flexible displays (32). The microsecond-range response time is two order of magnitudes faster than that of conventional LCs and related composite systems, such as the conventional polymer-dispersed LCs. At lower temperature $T_{PNI}-T=7.5$ °C, the nematogel's transmission is switched from 5% to 8% across the visible spectrum by the electric field of 3.7 V/μm, while it is switched from 5% to 22% by a stronger electric field of 7.8 V/μm. At an elevated temperature close to the paranematic-isotropic transition, $T_{PNI}-T=1.0$ °C, the electric field of 7.8 V/μm switches the nematogel's light transmission in a wide range from 20% to 61% (Fig. 3A,B). These characteristics can be further tuned, depending on the need, by using the infiltrating thermotropic LCs with different refractive indices pre-selected to optimize transparency in the field-on or field-off states and for different polarizations of incident light. To determine the response time, in accordance with the model detailed above, the transmission dependencies on time are fitted by exponentials $\eta_{trans} - \eta_{ref} = -A\exp(-t/\tau_{rising})$ and $\eta_{trans} - \eta_{ref} = B\exp(-t/\tau_{falling})$, where $\eta_{ref}$ is the reference transmission and $A$, $B$ are fitting coefficients. The sub-millisecond 0.1-0.35 ms response times are observed at all studied temperatures and voltages (Fig. 3A,B). Both the rising and falling times increase with temperature, which is mainly because the ratio $\gamma/K$ increases with temperature (Fig. 3C). The falling time remains almost unchanged at different voltages while the rising time decreases with the magnitude of the applied electric field, which is consistent with Eqs. (3) and (4) (Fig. 3D). The ultrafast response time arises from the constrains on LC molecule orientations in the vicinity of the bulk-distributed surfaces of the CNF network. Based on our model and Equation 3, taking $\gamma=0.045$ Pas of 5CB at 31.5 °C (33) and experimental $\tau_{falling}=117$ μs at electric field of 7.8 V/μm, we obtain the surface anchoring coefficient $W \approx 2.6\times10^{-5}$ J/m$^2$, which is close to the value deduced from the measurement of the threshold electric field, as discussed above. On the other hand, by using $W$ obtained from the independent measurements and estimates described above, we can use our model to describe/predict the experimental threshold field for electric switching of the nematogel composites, which again exhibits a good agreement.

**Thermally-tunable anisotropic mechanical properties**

From the standpoint of view of mechanical properties, the nematogel behaves as a liquid crystal gel with complex tensile elastic modulus modestly changing around the phase transition temperature of the 5CB infiltrating the cellulose network (Fig. 4A). The nematogel exhibits a predominantly elastic response with a storage modulus ($G'$) having a plateau in its temperature dependence at lower temperatures (Fig. 4A). The loss modulus ($G''$) also exhibits a plateau but starts to increase at temperatures comparable and above that of the nematic-isotropic transition, which is also accompanied by a slight decrease of $G'$ (Fig. 4A). Dynamic mechanical analysis measurements at low frequency reveal that the storage modulus stays roughly constant (~73.5 kPa) at temperatures below $T_{NI}$ and starts to gradually decrease above $T_{NI}$, while the loss modulus remains ~2.7 kPa below $T_{NI}$ and continuously increases by ~10% after increasing temperature above $T_{NI}$ of 5CB. These gradual changes of $G'$ and $G''$ are consistent with the "blurred" nematic-isotropic transition of the infiltrating 5CB fluid infiltrated into the cellulose gel network caused by paranematic ordering induced by cellulose nanofiber surfaces (consistent with the temperature behavior of optical characteristics discussed in Fig. 2). The nematogels exhibit anisotropic linear elasticity at high strains of 23% along the rubbing direction and 18% orthogonal to the rubbing direction (Fig. 4B). The temperature dependent behavior of mechanical properties of nematogels could be understood by the fact that they undergo structural transitions associated with decreasing surface coupling between LC molecules with CNFs skeleton above $T_{NI}$ and weakened hydrogen bonds between CNFs at higher temperature. Such unique viscoelastic properties of nematogels further enrich the behavior seen in other soft materials, such as flexible synthetic polymer gels (34), which respond linearly to stress with a constant $G'$, or biological polymer gels, which stiffen as they are strained. For example, the polyacrylamide hydrogels (5%) show a strain-independent shear storage modulus of ~110 Pa, smaller than that in our nematogels. The elasticity of collagen, fibrin, vimentin and neurofilaments are highly nonlinear, that is, the shear storage moduli are as small as 2-20 Pa at smaller strains but increase up to ten times at higher distortions (34). Nonlinear effects could also potentially arise in our nematogels when starting from polydomain samples, which could be caused by the interplay of healing of defects and grain boundaries between the domains as the distortion is increased, though this possibility will require further studies. As compared to our nematogels, the conventional cellulose hydrogels (at 0.8 wt.% of CNFs) crosslinked by diamines have a smaller shear storage modulus of ~1.2 kPa, a loss modulus of ~0.15 kPa and a Young's modulus of 2.5-3.5 kPa (35). Compared with other soft materials and cellulose hydrogels with weaker elasticity, the nematogel we fabricated shows rather high storage modulus of ~73.5 kPa at a low concentration of CNFs of less than 1 wt.%. The unique feature of the mechanical behavior of our nematogels is that the mechanical properties can be controlled by varying the degree of ordering of the host LC fluid through changing temperature, albeit within a relatively limited range (Fig. 4).

**Discussion and conclusions**

An interesting feature of ordered CNF-based nematogels is that the ordering direction of the nanofibers $\mathbf{N}_s$ defines the ordering of the small-molecule LC infiltrating it. Common applications of LCs in displays and electro-optic devices require treatment of confining surfaces with specialized alignment layers, which typically have to be rubbed to obtain monodomain LCs with the in-plane director alignment (9,28). In the case of nematogels, this control of the LC ordering direction is implemented through the shearing of the cellulose gel network, a technologically simpler process. Although a large number of flexible LC display approaches have been developed recently (36,37), they are commonly associated with technical challenges related to maintaining cell gap thickness and precluding LC fluid flow during the display flexing. These problems are solved naturally in the case of CNF-based nematogels that can be molded to yield desired geometric characteristics while sandwiched between flexible transparent electrodes. Moreover, apart from several notable exceptions (38,39), the fast switching characteristics demonstrated here for nematogels are hard to achieve within pristine LC systems, which may enable their uses in various electro-optical shutters and modulators. Furthermore, unlike many conventional gels, our nematogels are flexible rubbery materials which can be further processed and packaged by means of scalable technological processes such as lamination, rolling, *etc*. Although different LC transparency/scattering switching modes have been recently developed (32), our composite may allow for making such switching faster and (when implemented for thin cellulose-based nematogel films) at switching voltages comparable to those used in switching conventional LCs. For example, our LC nematogel films with micrometer-range thickness could be switched at electric fields above the critical value $E_c$=3.0 V/μm, which would translate to voltages ~10V, comparable to what is used for switching conventional nematic LCs (32).

To conclude, we have introduced a new breed of gels that combine hierarchical ordering of thermotropic LC molecules and that of CNFs within the gel network. These composite mesostructured soft solid materials combine properties such as mechanical flexibility, electric switching of transparency, strong temperature sensitivity, and ultrafast response to external stimuli like electric fields. We envisage a broad range of applications ranging from privacy and smart windows to electro-optic modulators and flexible displays. Our switchable nematogel composites will further expand the spectrum of emergent mesostructured cellulose-based materials with liquid crystalline ordering and diverse technological applications (40-43).

**Materials and Methods**

**Synthesis of CNFs**

Cellulose nanofibers with the dimensions of 7 nm×~400 nm (CNF1) were synthesized by TEMPO-mediated oxidation (44). Briefly, wood-cellulose based bleached coffee filter fragments (1 g) were suspended in water (100 mL) containing 16 mg of TEMPO (from Sigma-Aldrich) and 0.1 g of sodium bromide (from Sigma-Aldrich). The TEMPO-mediated oxidation was started by adding 2.5 mL of sodium hypochlorite solution (NaClO, 13% active chlorine, ACROS Organics), and was continued at room temperature by stirring at 500 rpm and occasional sonication. The pH was maintained at 10 by adding 0.5 M NaOH until no additional NaOH consumption was observed. The TEMPO-oxidized cellulose was thoroughly washed with water by filtration and mechanically homogenized as described below. CNFs with the dimension of 4.8 nm by 1.2 μm (CNF2) were synthesized using the same source material but by following the different procedures described elsewhere in literature (45). Briefly, wood-cellulose based bleached coffee filter (1 g) was suspended in 0.05 M sodium phosphate buffer (90 mL, pH 6.8) dissolving 16 mg of TEMPO and 1.13 g of 80% sodium chlorite in a flask. 455 μL of NaClO solution (13% active chlorine) was diluted ten times with the same 0.05 M buffer used as the oxidation medium and was added at one step to the flask. The flask was immediately stoppered, and the suspension was stirred at 500 rpm and 60 °C for 96 hours. After cooling the suspension to room temperature, the TEMPO-oxidized celluloses were thoroughly washed with water by filtration. TEMPO-pretreated celluloses were then mechanically blended by a food processor (from Oster), homogenized by a sonifier and filtered by a membrane filter.

**Preparation of a nematogel solid**
To obtain a nematogel with the desired geometric shape, 0.1-1.0 vol.% CNFs aqueous dispersion was poured into a mold, aligned by unidirectional shearing force and several drops of 1 M hydrochloric acid solution was added to prompt formation of a hydrogel within 2 hours. Typically, the shearing force was applied to the sample sandwiched between two glass plates separated by spacers (30 μm-1 mm). Then the top glass plate was sheared at a speed of 1 cm/s back and forth for 10 times. Multiple gelation time has been tested to ensure that two hours are long enough to prompt the gelation of CNFs solution. Then the hydrogel was immersed into isopropanol or other organic solvents for 2 days for solvent exchange. Finally the ensuing organogel was put into 5CB (Chengzhi Yonghua Display Materials Co. Ltd.) at 90 °C for 12 hours, with 5CB replacing isopropanol that was eventually evaporated. Because basic or hydrogen-bond-acceptor solvents which can destroy the hydrogen bonds between CNFs were avoided in this process, the ordered nanofiber network of the hydrogels was preserved in the organogel and nematogel throughout the solvent exchange process. The ensuing nematogel was cut into samples of desired dimensions (such as the thin films that were then used to obtain flexible nematogel devices) for further experimental studies reported here. The gel networks are birefringent as-fabricated well before 5CB or other thermotropic LC solvent is infiltrated. However, as we replace the conventional isotropic solvent with 5CB and lower temperature from the high values of 90 °C or higher, we see that the birefringence increases further, until eventually undergoing the transition to the bulk ordered state. We therefore conclude that both the nematic gel and the LC fluid infiltrating it are birefringent at room temperature and the overall birefringence of the composite results from the superposition of the two. To check that our approach can be extended to other material systems, in addition to infiltrating the cellulose gels with 5CB, similar studies have also been performed by using a commercial nematic mixture E7 (purchased from EM Chemicals), achieving similar behavior of nematogels but with higher order-disorder transition temperatures, which is due to the higher $T_{NI}$ of E7; only the experimental data obtained for 5CB-based nematogels are presented in this work for consistency.

**Preparation of a flexible LC panel**
The CNF aqueous dispersion was mixed with Mylar or Teflon spacers and sandwiched between two ITO coated plastic films. The plastic films were sheared unidirectionally so that an aligned CNF network could be achieved. Following this, the plastic cell was glued at corners by a UV-curable epoxy NOA-61 (from Norland Products) and placed under a UV lamp to cure it. The cell was placed in 5 wt.% acetic acid for 12 hours to make sure the CNF hydrogel network formation was complete. A weak acid, such as the acetic acid, was chosen for this procedure to avoid side effects associated with etching the ITO electrodes. Following this, water within the CNF hydrogel was exchanged by isopropanol for 12 hours. The nematogel was formed by replacing isopropanol with nematic LC 5CB (from Chengzhi Yonghua Display Materials Co. Ltd.).

**Imaging, electro-optical and mechanical characterization**

For polarizing and unpolarized-light brightfield optical microscopy observations, we used an Olympus BX-51 polarizing optical microscope equipped with 10 ×, 20 ×, and 50 × air objectives with a numerical aperture NA = 0.3-0.9 and a charge coupled device camera Spot 14.2 ColorMosaic (Diagnostic Instruments, Inc.). Transmission spectra were studied using a spectrometer USB2000-FLG (Ocean Optics) mounted on the microscope. Electric switching of nematogel was characterized using a data acquisition system SCC-68 (National Instruments Co.) controlled by a homemade software written in Labview (National Instruments Co.), a wideband power amplifier (Krohn Hite Co. Model 7600) and a Si-amplified photodetector PDA100A (Thorlabs Inc.). The nematogel has been switched thousands of times during the measurement and no detectable change on switching properties has been observed. The composites were also switched months apart in time, without exhibiting difference in their performance. Transmission electron microscopy (TEM) images were obtained using a CM100 microscope (FEI Philips). The CNFs samples were negatively stained by phosphotungstic acid to increase the contrast of images: 2 μL of the sample is deposited on the formvar coated copper grid, allowed to settle for drying, and then dipped into the stain solution (2 wt.% phosphotungstic acid). SEM images of CNF aerogels were obtained using a Carl Zeiss EVO MA 10 system: fresh surfaces of the tearing CNF aerogels were sputtered with a thin layer of gold and observed under TEM at a low voltage of 5kV to avoid the distortion of aerogel samples. A Q800 dynamic mechanical analyzer (TA Instruments) was used to probe mechanical properties and determine the modulus behavior *versu*s temperature. Nematogels with different physical dimensions and CNF concentrations were also fabricated and each sample was measured 3-4 times. The mechanical properties shown in Fig. 4 were measured with nematogel samples with 0.25 vol.% CNFs cut into rectangular strips of 20 mm×6 mm×1 mm. During the mechanical property characterization, the temperature was ramped from 25 °C to 50 °C at a rate of 3 °C min$^{-1}$, using a frequency of 1 Hz and oscillatory strain of 2.0 %. Photographs of nematogel samples and flexible devices based on them were made using a Nikon D300 camera.

**Acknowledgments:** We thank B. Senyuk, L. Jiang and Y. Yuan for discussions and D. Zhang, D. Rudman, H. B. Song and M. McBride for technique assistance.

**Funding:** This research was supported by the U.S. Department of Energy, ARPA-E award DE-AR0000743.

**Author contributions**: I. I. S conceived and supervised the project. Q. L. carried out all fabrication and characterization of the cellulose nematogel. I. I. S. and Q. L. discussed the results and wrote the paper.

**Competing interests:** The authors declare that they have no competing interests.

**Data and materials availability:** All data needed to evaluate the conclusions in the paper are present in the paper. Additional data available from authors upon request.

**Figures**

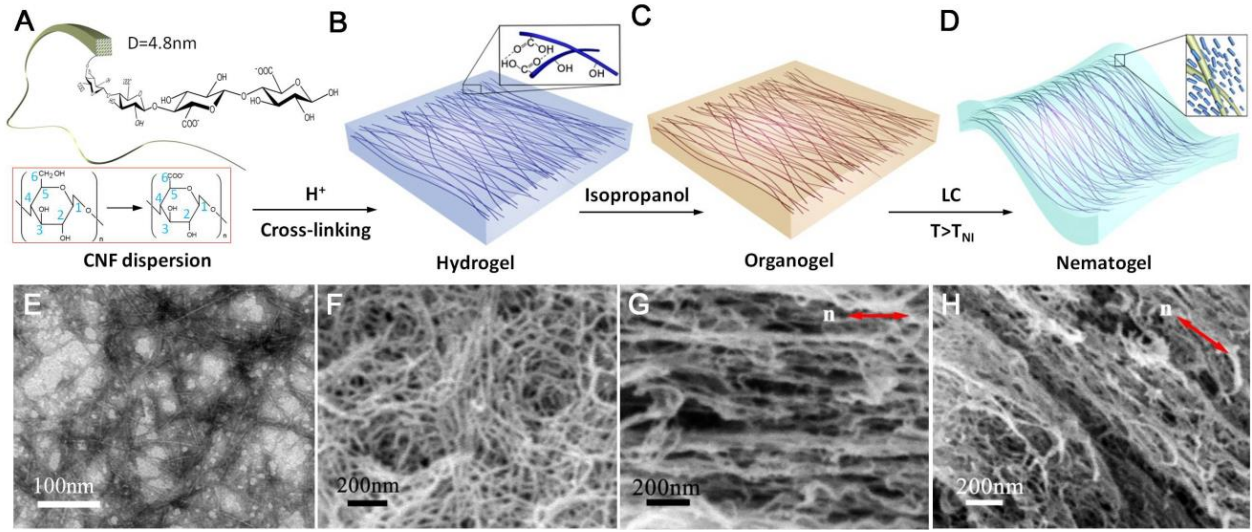

**Fig. 1. Fabrication of cellulose-based nematogels.** (**A**) Structure of CNFs initially obtained in an aqueous dispersion. Inset: chemical structure of cellulose before (left) and after (right) the oxidation process. (**B**) A hydrogel with inter-linked CNFs formed upon cross-linking of the individual fibers into a network, which can be aligned by unidirectional shearing before cross-linking. (**C**) Organogel obtained via the replacement of water with isopropanol or other organic solvents. (**D**) Nematogel obtained via the substitution of the organic solvent with a LC such as 5CB in nematic phase at room temperature. (**E**) TEM image of CNF2 negatively stained by 2 wt.% phosphotungstic acid solution. (**F**-**H**) SEM images of (**F**) unaligned CNF2 aerogel and (**G**,**H**) aligned CNF2 aerogels with (**G**) 0.12 vol.% and (**H**) 0.6 vol.% of CNFs, coated with a thin layer of gold and observed at a low voltage of 5 kV.

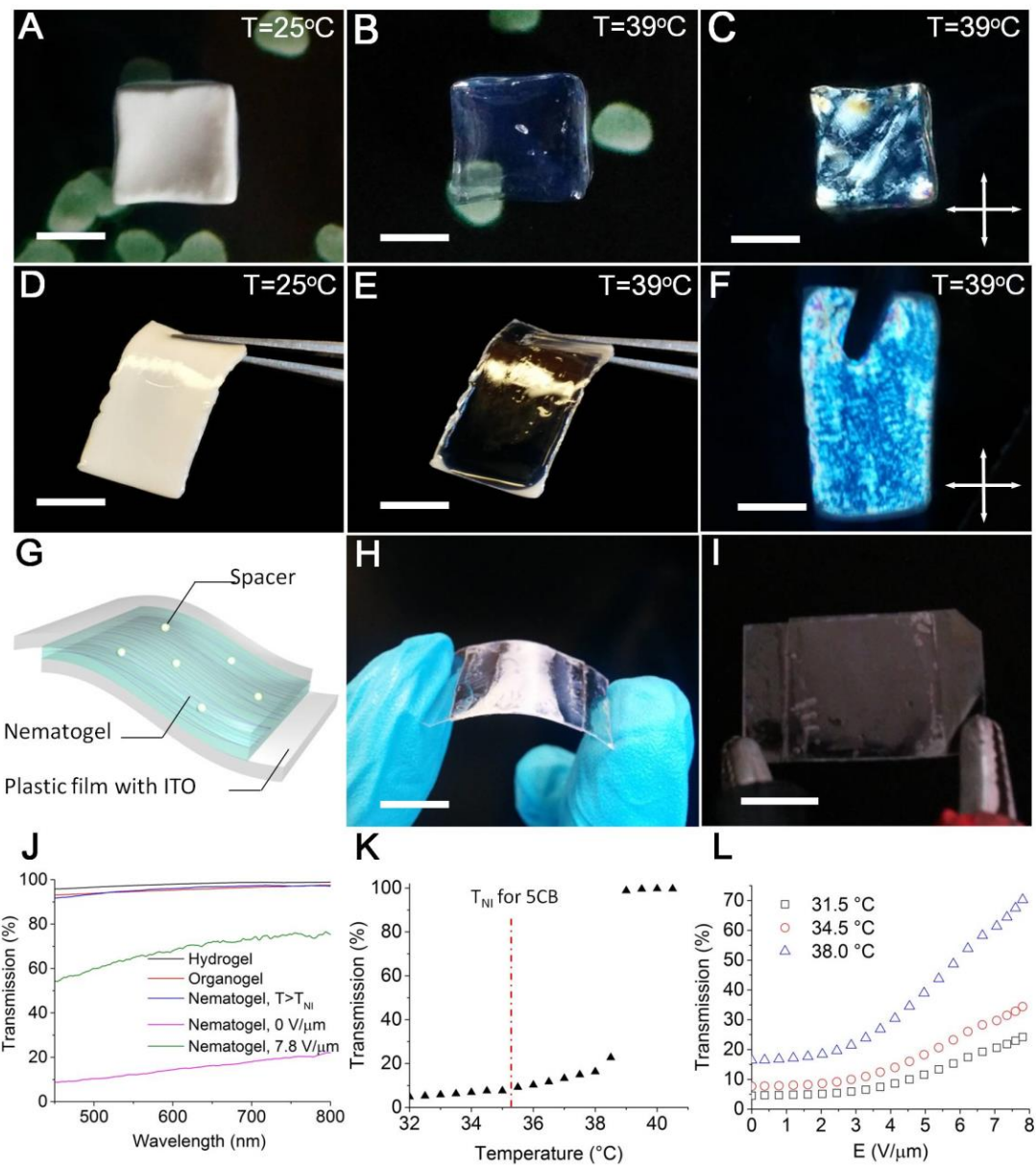

**Fig. 2. Optical properties of nematogels and their use in fabricating stimuli-responsive devices.** Photographs of nematogel films at different temperatures (indicated in the top-right corners) corresponding to (**A,D**) nematic (**A**:CNF1; **D**: CNF2) and (**B,C,E,F**) isotropic phase of the 5CB infiltrating the cellulose gel network (**B,C**:CNF1; **E,F**: CNF2). The photographs (**C,F**) were obtained for nematogel samples placed between two crossed polarizers. (**G**) Schematic of a flexible nematogel LC cell fabricated by confining the nematogel film between two plastic films with transparent ITO electrodes facing inwards. Photographs of the flexible nematogel cell (CNF2-5CB) in (**H**) nematic phase and (**I**) isotropic phase, with the electrodes used to apply fields for electro-optic characterization like that shown in (**J,L**). (**J**) Transmission of 1 mm-thick hydrogel, organogel and nematogel and 30 μm-thick CNF2-5CB nematogel with and without electric field, as indicated in the legend. (**K**) Transmission of CNF2-5CB nematogel *vs.* temperature, showing that the paranematic to isotropic transition takes place

at $T_{PNI} \approx 38.5°C$; for reference, the temperature of nematic-isotropic transition for a pristine 5CB $T_{NI} \approx 35.3°C$ is indicated using a dashed red vertical line. (**L**) Transmission of CNF2-5CB nematogel *vs.* voltage at different temperatures, showing that the critical field $E_c \approx 3.0$ V/μm varies only weakly with temperature. All scale bars are 1 cm.

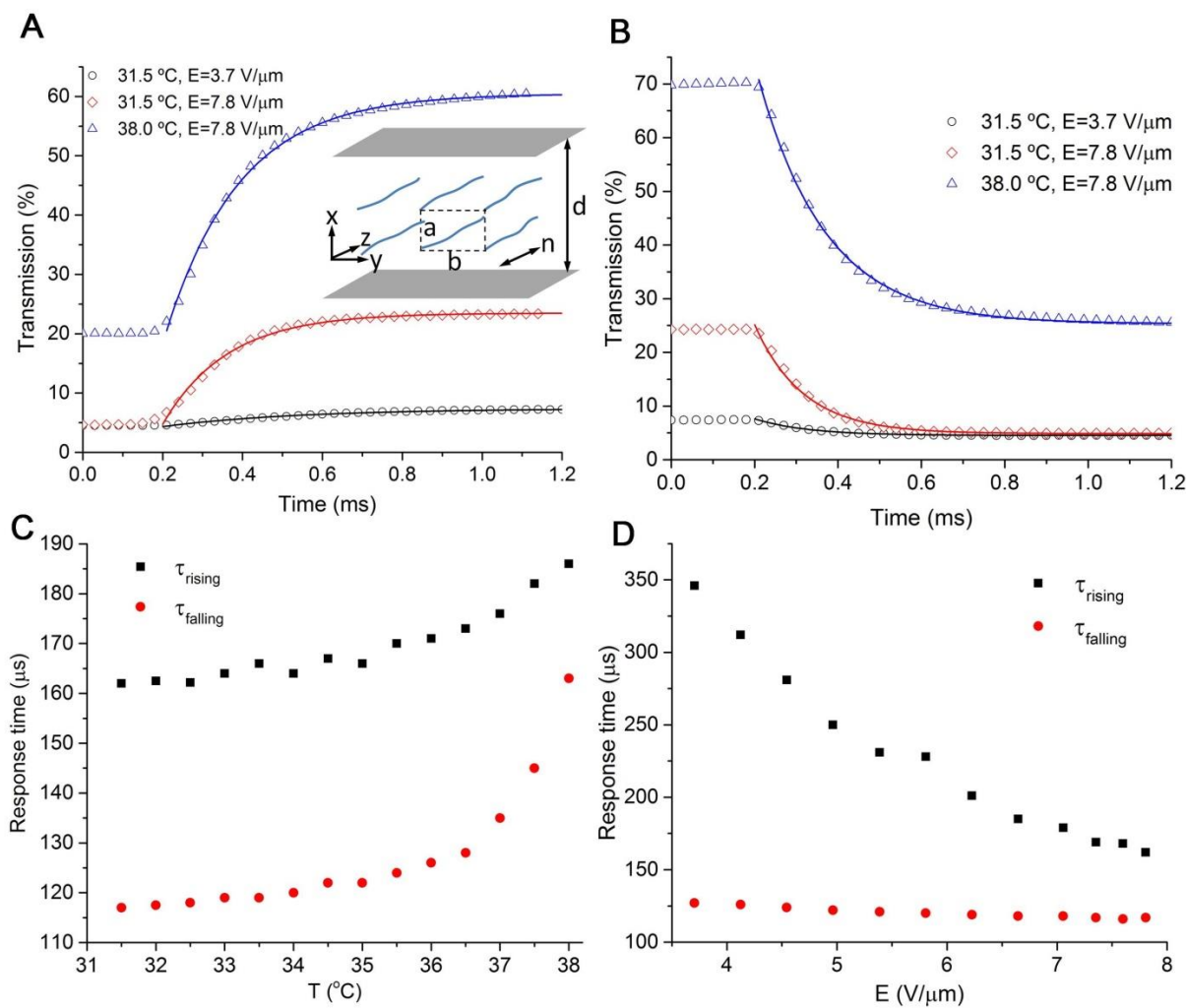

**Fig. 3. Characterization of response times associated with electric switching of nematogels.** (**A**,**B**) Transmission *vs.* time curves used to characterize (**A**) rising time and (**B**) falling time of CNF2-5CB nematogel at different temperatures and applied voltages indicated in the legends. The inset in (**A**) shows the coordinate system and the physical model of the nematogel cell of thickness *d* and with a CNF-compartmentalized LC domain characterized by geometric parameters *a* and *b*; the uniform director at no fields and the CNF fibers are shown co-aligned. The experimental data (scatter symbols) are fitted by the corresponding results emerging from the physical model (solid lines). (**C**) Response times of CNF2-5CB nematogel *vs.* temperature, where $\tau_{rising}$ and $\tau_{falling}$ were measured using the field of 7.8 V/μm. (**D**) Response time *vs.* voltage characterized at 31.5 °C.

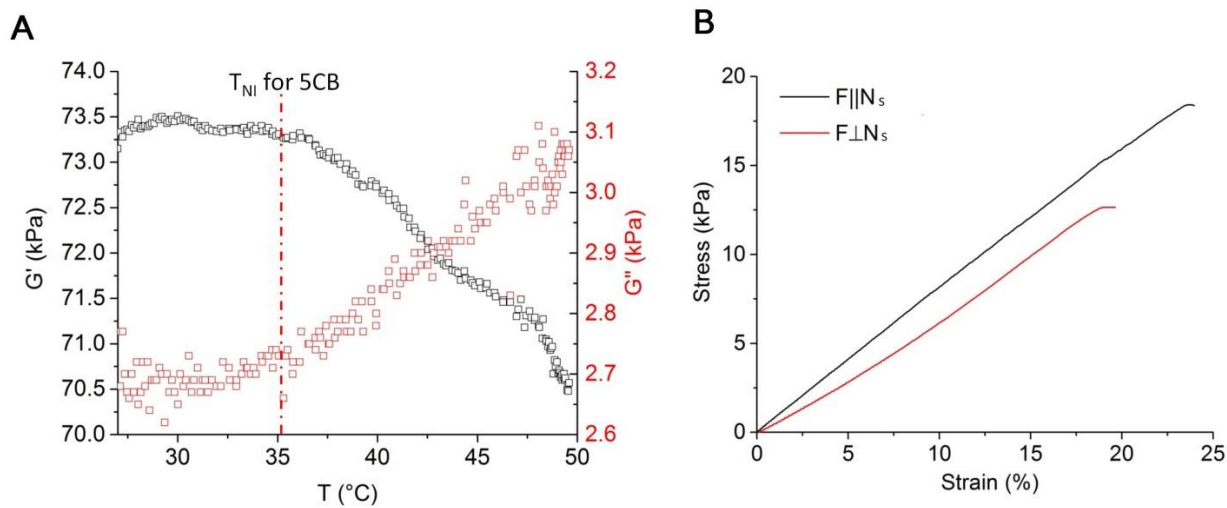

**Fig. 4. Characterization of mechanical properties of nematogels.** (**A**) Tensile elastic modulus versus temperature of CNF2-5CB nematogel. The storage modulus $G'$ (black open squares) and the loss modulus $G''$ (red open squares) are plotted using the left and right vertical axes, respectively. (**B**) Stain-stress relation along and perpendicular to $\mathbf{N}_s$.